\documentclass[12pt,a4paper]{article}

\usepackage{a4wide}
\usepackage{amsmath}
\usepackage{bm}
\usepackage{amssymb}
\usepackage{hyperref}
\usepackage{epic}

\newcommand{\cN}{{\mathcal N}}

\newcommand{\CA}{C_A}
\newcommand{\zt}{\zeta_3}

\newcommand{\Dim}{\mathcal{D}}
\newcommand{\mksi}{\alpha}
\newcommand{\ep}{{\epsilon}}
\newcommand{\DR}{\mathrm{DR}}

\makeatletter
\def\mr@ignsp#1 {\ifx\:#1\@empty\else #1\expandafter\mr@ignsp\fi}%
\newcommand{\multiref}[1]{\begingroup
\xdef\mr@no@sparg{\expandafter\mr@ignsp#1 \: }%
\def\mr@comma{}%
\@for\mr@refs:=\mr@no@sparg\do{\mr@comma\def\mr@comma{,}\ref{\mr@refs}}%
\endgroup}
\makeatother

\begin{document}
\thispagestyle{empty}


\begin{center}
{\Large{\bf
Three-loop renormalization of the $\cN=1$, $\cN=2$, $\cN=4$\\[4mm] supersymmetric Yang-Mills theories
}}
\vspace{15mm}

{\sc
V.~N.~Velizhanin}\\[5mm]

{\it Theoretical Physics Department\\
Petersburg Nuclear Physics Institute\\
Orlova Roscha, Gatchina\\
188300 St.~Petersburg, Russia}\\[5mm]

\textbf{Abstract}\\[2mm]
\end{center}

\noindent{We calculate the renormalization constants
of the $\cN=1$, $\cN=2$, $\cN=4$ supersymmetric Yang-Mills theories in an arbitrary covariant gauge
in the dimensional reduction scheme up to three loops.
We have found, that the beta-functions for $\cN=1$ and $\cN=4$ SYM theories
are the same from
the different triple vertices.
This means that the dimensional reduction scheme works correctly in these models up to third
order of perturbative theory.}
\newpage

\setcounter{page}{1}

Regularization by dimensional reduction was proposed by Siegel~\cite{Siegel:1979wq}
for calculations in the supersymmetric theories.
It has a simple explanation as dimensional reduction
from the higher dimensions~\cite{Siegel:1979wq,Capper:1979ns}.
For example, if we start from $\cN=1$ supersymmetric Yang-Mills (SYM) theory in 10 dimensions,
where the number of bosonic degrees of freedom is equal to fermionic degrees of freedom, this number should be the same
when we go to $\cN=4$ SYM theory in 4 dimensions.
In this way one obtains, that the number of the scalar
particles should be $6+2\,\ep$ with $2\,\ep$ additional  scalars , where
$\ep$ is the parameter of the dimensional regularization,
the dimension of the space-time being $d=4-2\,\ep$.

However, as was pointed out by Siegel
himself~\cite{Siegel:1980qs} and then studied
in Refs.~\cite{Capper:1979ns,Avdeev:1981vf,Avdeev:1982np,Avdeev:1982xy}
the dimensional reduction scheme has some inner problems
and should be violated at higher-loop orders. In particular for the $\cN=4$ SYM theory
for the propagator type diagrams DR-scheme should works at least up to ten loops and for
the triple vertices at least up to eight loops~\cite{Avdeev:1981vf}.
However, later in Ref.~\cite{Avdeev:1982np} were presented the results for the three-loop $\beta$-functions
in $\cN=1$, $\cN=2$ and $\cN=4$ SYM theories in $\Dim=4$ dimension
or, equivalently, for $\cN=1$ SYM theories in $\Dim=4$, $\Dim=6$ and $\Dim=10$ dimensions correspondingly,
obtained from the different vertices, namely, from the
fermion-fermion-vector and fermion-fermion-scalar vertices.
It was claimed that the three-loop $\beta$-functions are different from these vertices.
This result means that the gauge and Yukawa coupling constants are renormalized in different ways
so that the DR-scheme violates supersymmetry and
does not work already at this level for any dimension $\Dim$~\cite{Avdeev:1982np,Avdeev:1982xy}.

To study this problem for the future four-loop computations,
we have repeated these calculations (but in the
arbitrary covariant gauge) and have found,
that indeed, in general for the arbitrary dimension $\Dim$ the $\beta$-functions are different
from these vertices, but for $\Dim=4$\footnote{A fact,
that the result of Ref.~\cite{Avdeev:1982np} is incorrect
for $\cN=1$ SYM theories in $\Dim=4$ dimension was pointed out firstly in
Ref.~\cite{Harlander:2006xq}.} and $\Dim=10$
($\cN=1$ and $\cN=4$ SYM theories in $\Dim=4$) these $\beta$-functions are the same
on the contrary to the statement of Refs.~\cite{Avdeev:1982np,Avdeev:1982xy}.
This result allows to hope, that the limitations from Table 1 in Ref.~\cite{Avdeev:1982np}
are correct and it is possible to use $\DR$-scheme beyond three loops.

Renormalization constants within MS-like schemes do not
depend on dimensional parameters (masses, momenta)~\cite{Collins:1974bg} and
have the following structure:
\begin{eqnarray}
\label{eq:5}
Z_\Gamma\!\left(\frac{1}{\epsilon},\alpha,g^2\right)=
1+\sum^\infty_{n=1}c_\Gamma^{(n)}\!\!\left(\alpha,g^2\right)\epsilon^{-n},
\end{eqnarray}
where $\alpha$ is the gauge fixing parameter.
The renormalization constants define corresponding anomalous dimensions:
\begin{eqnarray}
\label{defga}
\gamma_\Gamma(\alpha,g^2)=
g^2\frac{\partial}{\partial g^2}\ c^{(1)}_\Gamma\!(\alpha,g^2).
\end{eqnarray}

For the calculation of the renormalization constants,
following of Ref.~\cite{Larin:1993tp}
(see also Refs.~\cite{Tarasov:1976ef,Vladimirov:1979zm,Tarasov:1980kx}),
we use the multiplicative renormalizability of Green functions.
The renormalization constants $Z_\Gamma$ relate the dimensionally regularized
one-particle-irreducible
Green function with renormalized one as:
\begin{eqnarray}
\label{multren}
\Gamma_{\mathrm{Renormalized}}\left(\frac{Q^2}{\mu^2},\alpha,g^2\right)&=&\lim_{\epsilon \rightarrow 0}
Z_\Gamma\left(\frac{1}{\epsilon},\alpha,g^2\right)
\Gamma_{\mathrm{Bare}}\left(Q^2,\alpha_{\mathrm{B}},g^2_\mathrm{B},\epsilon\right),
\end{eqnarray}
where $g^2_{\mathrm{B}}$ and $\alpha_{\mathrm{B}}$ are the bare charge and gauge fixing parameter
correspondingly with
\begin{eqnarray}
g^2_{\mathrm{B}}&=&
\mu^{2\epsilon}\left[g^2+\sum_{n=1}^{\infty}a^{(n)}\!\left(g^2\right)\epsilon^{-n}\right],\label{gbex}\\[5mm]
\alpha_{\mathrm{B}}&=&\alpha Z_3\,,\qquad Z_3=Z_g\,.
\end{eqnarray}
The bare charge $g^2_{\mathrm{B}}$ is to be constructed from appropriate $Z_i$. In general
for the triple vertices we have
\begin{equation}
g^2_{\mathrm{B}}=\mu^{2\epsilon} g^2 Z_{jjk}^2 Z_j^{-2} Z_k^{-1}\,,\label{gbz}
\end{equation}
where $Z_{jjk}$ and $Z_j$ are the renormalization constants for the triple vertices and
the wave functions correspondingly.
From eqs.~(\ref{gbex}) and~(\ref{gbz}) one obtains the charge renormalization $\beta$-function as
\begin{eqnarray}
\beta_{jjk}\left(g^2\right)&\equiv&\left(g^2\frac{\partial}{\partial g^2}-1\right)
a^{(1)}_{jjk}\!\left(g^2\right)=
g^2 \left[2\,\gamma_{jjk}\!\left(\alpha,g^2\right)-2\,\gamma_{j}\!\left(\alpha,g^2\right)
-\gamma_{k}\!\left(\alpha,g^2\right)\right].\label{beta}
\end{eqnarray}

The calculation of the renormalization constants within MS-like scheme can
be reduced to the calculation only
of massless propagator type diagrams by means of the method of
infrared rearrangement~\cite{Vladimirov:1979zm}.
In the case of the gauge (fermion-fermion-vector, scalar-scalar-vector, ghost-ghost-vector) or Yukawa (fermion-fermion-scalar) vertices it means that we can nullify the momentum of the external vector or scalar fields, correspondingly,
reducing the calculation of the $Z_{jjk}$ to the propagator type diagrams.

To find the renormalization constants we compute with the FORM~\cite{Vermaseren:2000nd}
package MINCER~\cite{Gorishnii:1989gt}
the unrenormalized three-loop
one-particle-irreducible fermion-fermion-vector,
scalar-scalar-vector, ghost-ghost-vector, fermion-fermion-scalar vertices
and inverted fermion, scalar, ghost and vector propagators.
Having the two-loop
bare charge we determine the necessary three-loop constants
$Z_{jjk}$ and $Z_{j}$ from the requirement that the poles
in $\ep$ cancel in the r.h.s. of Eq.~(\ref{multren}).
We use a program DIANA~\cite{Tentyukov:1999is},
which call QGRAF~\cite{Nogueira:1991ex} to generate all diagrams.
The computations were performed using the FORM~\cite{Vermaseren:2000nd}, the FORM package COLOR~\cite{vanRitbergen:1998pn} for evaluation of the color traces and with Feynmans rules from Refs.~\cite{Gliozzi:1976qd,Avdeev:1980bh}
with the arbitrary gauge fixing parameter $\alpha$, i.e. the propagator of the vector field is
$(g_{\mu\nu}-(1-\alpha)q_\mu q_\nu/q^2)/q^2$.

Substituting the obtained $\gamma$-functions into Eq.~(\ref{beta})
we have found from the fermion-fermion-vector,
scalar-scalar-vector, ghost-ghost-vector vertices the following $\beta$-function
($\CA$ is the quadratic Casimir invariant, $C_A=N_c$ for $SU(N_c)$):
\begin{equation}\label{betag}
\beta^{3-\mathrm{loop}}(a)\ =\
\frac12\,(\Dim-10)\, \CA\, a^2 \bigg[
1
- (\Dim-6)\,  \CA\, a
+ \frac{7}{4}\; (\Dim-6)^2\; \CA^2\, a^2\bigg],\quad a=\frac{g^2}{(4\pi)^2}
\end{equation}
in accordance with the previous calculations~\cite{Avdeev:1981ew},
while from the fermion-fermion-scalar vertex we have received
($\hat{d}_{44}$ becomes the quartic Casimir invariant~${d}_{44}=N_c^2(N_c^2+36)/24$~\cite{vanRitbergen:1998pn}, after contraction $(\hat{d}_{44}f^{abc})$ with $f^{abc}$)
\begin{eqnarray}\label{betay}
\beta_{ffs}^{3-\mathrm{loop}}(a)&=&
\frac12\,(\Dim-10)\, a^2\, \Bigg[
\CA
- (\Dim-6)\,  \CA^2\, a\nonumber\\
&+&
\bigg\{\left(
\frac{1}{12}  \left(\Dim^2-4 \Dim+84\right)
+ (\Dim-4) (2 \Dim-15)\, \zt\right)\CA^3\nonumber\\
&+&
(\Dim-4) \Big(4\, (\Dim-3) -24\,  (2 \Dim-15)\, \zt\Big)\hat{d}_{44}\bigg\}\;a^2\;\Bigg],
\end{eqnarray}
which is different with compare to the result from Ref.~\cite{Avdeev:1982np}.
For $\Dim=10$ and $\Dim=4$ the $\beta$-functions (\ref{betag}) and (\ref{betay}) are the same,
on the contrary to the result of Ref.~\cite{Avdeev:1982np}.
For $\Dim=6$ the $\beta$-function~(\ref{betay}) from the Yukawa vertex is nonzero at three loops,
as in Ref.~\cite{Avdeev:1982np} but with different coefficients.
The equivalence of the three-loop $\beta$-functions from the gauge and fermion-fermion-scalar vertices
for the $\cN=1$ SYM theory in $\Dim=4$ was obtained already in Ref.~\cite{Harlander:2006xq}, that has
allowed to find the four-loop $\beta$-function in this model from the corresponding result
in QCD~\cite{vanRitbergen:1997va}.

\subsubsection*{Note added~\footnote{February 19, 2026; see also the Erratum~\cite{Velizhanin:2026kpj} to the journal version of this paper.}~\footnote{When this mismatch was corrected, we received information that the zero three-loop coefficient in the Yukawa beta function for $\cN=2$ SYM theory was obtained in Ref.~\cite{Chakraborty:2026hdv}.}}
In order to keep the dimension $\Dim$ arbitrary, we followed Ref.~\cite{Avdeev:1980bh} and imposed certain relations among the matrices appearing in the Yukawa vertices ($\alpha^r_{mn}$ and $\beta^r_{mn}$ in the notations of Refs.~\cite{Gliozzi:1976qd, Avdeev:1980bh}). In particular, we used the relation given in Eq.~(2) of Ref.~\cite{Avdeev:1980bh},
\begin{equation}
{\mathrm{tr}}[\alpha^r\beta^t]=0\,.\label{tralbe}
\end{equation}
However, this relation is valid only for $\cN=4$ SYM theory. For $\cN=1$ and $\cN=2$ SYM theories, this trace is generally nonzero, since $\alpha^r$ and $\beta^r$ are not matrices in these models\footnote{Other relations for $\alpha^r$ and $\beta^r$ from Ref.~\cite{Avdeev:1980bh} can still be used to obtain results keeping $\Dim$ unspecified.}. Therefore, we recomputed the renormalization of the Yukawa vertex while treating ${\mathrm{tr}}[\alpha^r\beta^t]$ as an independent quantity, obtaining
\begin{equation}\label{addcorr}
\frac{a^3}{3\; \ep} \Big(\CA^3 (4 - 3\, \zt) + 12\,(1 + 6\, \zt)\,\hat{d}_{44}\Big)\,{\mathrm{tr}}[\alpha^r\beta^t]\,\beta^t_{mn}\,.
\end{equation}
This additional term vanishes in $\cN=4$ SYM theory due to Eq.~(\ref{tralbe}), and in $\cN=1$ SYM theory due to the presence of an $\ep$-scalar inside the loop\footnote{The contribution of the $\ep$-scalar inside the loop is proportional to $\ep$, whereas the result in Eq.~(\ref{addcorr}) is proportional to $1/\ep$, yielding a finite expression.}. In contrast, for $\cN=2$ SYM theory this term enters the renormalization of the Yukawa vertex. It exactly cancels the last two lines of Eq.~(\ref{betay}), leading to a vanishing three-loop beta function $\beta_{ffs}^{3-\mathrm{loop}}$ in $\cN=2$ SYM theory.
Since this contribution is zero for both $\cN=4$ and $\cN=1$ SYM theories, the corresponding coefficient can be written as 
$c(\Dim-10)(\Dim-4)$, with $c=-1/4$ fixed by the $\cN=2$ SYM case. As a result, instead of Eq.~(\ref{betay}) we obtain
\begin{eqnarray}\label{betayr}
\beta_{ffs}^{3-\mathrm{loop}}(a)&=&
\frac12\,(\Dim-10)\, a^2\, \Bigg[
\CA
- (\Dim-6)\,  \CA^2\, a\nonumber\\
&+&
(\Dim-6)\bigg\{\frac{\Dim-46}{12}\,\CA^3+2\,(\Dim-4)\Big(\zt+2\,(1-12\,\zt)\,\hat{d}_{44}\Big)
\bigg\}\;a^2\;\Bigg].
\end{eqnarray}
So, we have found, that the gauge and Yukawa couplings are renormalized in the same way for all $\cN=1$, $\cN=2$ and $\cN=4$ SYM theory (or $\cN=1$ SYM theory in $\Dim=4$, $\Dim=6$ and $\Dim=10$). 
Then, the $\DR$-scheme preserves supersymmetry and works correctly in these models up to three loops.

In Appendix we give the renormalization constants for all fields in $\cN=1$, $\cN=2$ and $\cN=4$
SYM theories in $\Dim=4$. In general the renormalization constants for all vertices can be found from
Eqs.~(\ref{beta}) and~(\ref{betag}) for the triple vertices and the analogous equations for the
quarter vertices. For the fermion-fermion-vector, scalar-scalar-vector, ghost-ghost-vector vertices
we have found by the direct calculations the correctness of obtained renormalization constants, while for
the fermion-fermion-scalar vertex one should use Eq.~(\ref{betayr}) instead of Eq.~(\ref{betag}).

To conclude, we note that the obtained renormalization constants were used for the full direct calculation
of the four-loop anomalous dimension of Konishi operator in $\cN=4$ SYM theory~\cite{Velizhanin:2008jd}.
The result of this calculation 
coincides with the results of the corresponding 
superfield~\cite{Fiamberti:2007rj}
and superstring~\cite{Bajnok:2008bm} calculations.

\subsection*{Acknowledgements}
We would like to thank L.N. Lipatov, A.I. Onishchenko and A.A. Vladimirov for useful discussions.
This work is supported by
RFBR grants 07-02-00902-a, RSGSS-3628.2008.2.
We thank the Max Planck Institute for Gravitational Physics
at Potsdam for hospitality while working on parts of this
project.


\section*{Appendix: Renormalization constants}

In this Appendix we give explicit expressions for the renormalization
constants of the vector, fermion, scalar (pseudoscalar) and ghost fields
in $\mathcal{N}=4$, $\mathcal{N}=2$ and $\mathcal{N}=1$ SYM theories
up to three loops in the arbitrary covariant gauge. The renormalization constants for all
vertices can be easily found from the $\beta$-functions
~(\ref{beta}),~(\ref{betag}) and~(\ref{betayr})
and their higher poles.
\begin{eqnarray}
Z_g^{\mathcal{N}=4}&=&
1
-\frac{\mksi+3}{2 \ep}\CA a
+\left(\frac{2\mksi^2+9 \mksi+9}{8 \ep^2}
-\frac{2\mksi^2+11 \mksi-21}{16 \ep}\right) \CA^2 a^2
\nonumber\\
&+&\left(-\frac{2 \mksi^3+12 \mksi^2+21 \mksi+9}{16 \ep^3}
+\frac{14 \mksi^3+96 \mksi^2+27\mksi-189}{96 \ep^2}
\right.\nonumber\\
&&\left.
\quad-\frac{7 \mksi^3+33 \mksi^2-97 \mksi+175}{ 96\ep}
-\frac{\mksi^2+4 \mksi+79}{ 16\ep}\; \zt\right)\CA^3 a^3,
\\
Z_f^{\mathcal{N}=4}&=&
1
-\frac{\mksi+3}{\ep}\,\CA \,a
+\left(3\,\frac{\mksi^2+5 \mksi+6}{4\ep^2}
-\frac{\mksi^2+8 \mksi-33}{8\ep}\right)  \CA^2 \,a^2
\nonumber\\
&+&\left(
-\frac{4\mksi^3+27 \mksi^2+57 \mksi+36}{8\ep^3}
+\frac{2 \mksi^3+17\mksi^2-12 \mksi-99}{8 \ep^2}
\right.\nonumber\\
&&\left.
\quad-\frac{10 \mksi^3+39 \mksi^2-255\mksi+1014}{96\ep}
-\frac{\mksi^2+2\mksi+69}{8\ep}\,\zt\right) \CA^3 a^3,
\\
Z_s^{\mathcal{N}=4}&=&
1
-\frac{\mksi+1}{\ep}\,\CA\,a
+\left(\frac{3 \mksi^2+7 \mksi+2}{4\ep^2}
-\frac{\mksi^2+8 \mksi-5}{8\ep}\right)\,\CA^2 \,a^2
\nonumber\\
&+&\left(
-\frac{12 \mksi^3+45 \mksi^2+39\mksi+4}{24 \ep^3}
+\frac{2 \mksi^3+15 \mksi^2-5}{8 \ep^2}
\right.\nonumber\\
&&\left.
\quad-\frac{10\mksi^3+39 \mksi^2-247 \mksi+14}{96\ep}
-\frac{\mksi^2+35}{8\ep}\,\zt\right)\, \CA^3\,a^3,
\\
Z_{gh}^{\mathcal{N}=4}&=&
1
-\frac{\mksi-3}{4 \ep}\,\CA\,a
+\left(\frac{\mksi-21}{32 \ep}
+3\,\frac{\mksi^2+3}{32 \ep^2}\right)\, \CA^2\, a^2
\nonumber\\
&+&\left(
-\frac{5 \mksi^3+9 \mksi^2+3 \mksi-9}{128 \ep^3}
+\frac{8 \mksi^3+39 \mksi^2-18 \mksi-189}{384 \ep^2}
\right.\nonumber\\
&&\left.
\quad+\frac{-3 \mksi^3-6 \mksi^2+144 \mksi+175}{192\ep}
+\frac{\mksi^2+4 \mksi+79}{32\ep}\,\zt\right)\,\CA^3\, a^3;
\end{eqnarray}

\begin{eqnarray}
Z_g^{\mathcal{N}=2}&=&
1
-\frac{\mksi-1}{2 \ep}\,\CA\,a
+\left(\frac{2\mksi^2+\mksi-3}{8 \ep^2}
-\frac{2\mksi^2+11 \mksi+7}{16 \ep}\right)\,\CA^2 \,a^2
\nonumber\\
&+&\left(
-\frac{2 \mksi^3+4 \mksi^2+\mksi-7}{16 \ep^3}
+\frac{14 \mksi^3+72 \mksi^2+79\mksi+35}{96 \ep^2}
\right.\nonumber\\
&&\left.
\quad-\frac{7\mksi^3+33 \mksi^2+43 \mksi-273}{96\ep}
-\frac{\mksi^2+4 \mksi+39}{16\ep}\,\zt\right)\,\CA^3\, a^3,
\\
Z_f^{\mathcal{N}=2}&=&
1
-\frac{\mksi+1}{\ep}\,\CA\,a
+\left(\frac{3 \mksi^2+7 \mksi+6}{4\ep^2}
-\frac{\mksi^2+8 \mksi+7}{8\ep}\right)\,\CA^2\,a^2
\nonumber\\
&+&\left(-\frac{4 \mksi^3+15\mksi^2+25 \mksi+20}{8 \ep^3}
+\frac{6\mksi^3+45 \mksi^2+80 \mksi+49}{24\ep^2}
\right.\nonumber\\
&&\left.
\quad-\frac{10 \mksi^3+39 \mksi^2+21\mksi-410}{96\ep}
-\frac{\mksi^2+2\mksi+29}{8\ep}\,\zt\right)\, \CA^3\,a^3,
\\
Z_s^{\mathcal{N}=2}&=&
1
-\frac{\mksi-1}{\ep}\,\CA\,a
+\left(\frac{3\mksi^2-\mksi-2}{4 \ep^2}
-\frac{\mksi^2+8 \mksi+3}{8 \ep}\right)\,\CA^2\,a^2
\nonumber\\
&+&\left(\frac{-4 \mksi^3-3 \mksi^2+3\mksi+4}{8 \ep^3}
+\frac{6 \mksi^3+39\mksi^2+20 \mksi+3}{24\ep^2}
\right.\nonumber\\
&&\left.
\quad-\frac{10 \mksi^3+39 \mksi^2+29\mksi-226}{96\ep}
-\frac{\mksi^2+19}{8\ep}\,\zt\right)\, \CA^3\, a^3,
\\
Z_{gh}^{\mathcal{N}=2}&=&
1
-\frac{\mksi-3}{4 \ep}\,\CA \,a
+\left(\frac{\mksi+7}{32 \ep}
+3\,\frac{\mksi^2-5}{32 \ep^2}\right)\, \CA^2\,a^2
\nonumber\\
&+&\left(\frac{-5 \mksi^3-9 \mksi^2+5 \mksi+65}{128\ep^3}
+\frac{8 \mksi^3+39 \mksi^2+2 \mksi-49}{384 \ep^2}
\right.\nonumber\\
&&\left.
\quad-\frac{\mksi^3+2 \mksi^2-16 \mksi+91}{64\ep}
+\frac{\mksi^2+4 \mksi+39}{32\ep} \,\zt\right)\, \CA^3\,a^3;
\end{eqnarray}

\begin{eqnarray}
Z_g^{\mathcal{N}=1}&=&
1
-\frac{\mksi-3}{2 \ep}\,\CA\,a
+\left(\frac{2\mksi^2-3 \mksi-9}{8 \ep^2}
-\frac{2\mksi^2+11 \mksi-27}{16 \ep}\right)\,\CA^2 \, a^2
\nonumber\\
&+&\left(\frac{-2 \mksi^3+9 \mksi+27}{16 \ep^3}
+\frac{14 \mksi^3+60 \mksi^2-39\mksi-369}{96 \ep^2}
\right.\nonumber\\
&&\left.
\quad-\frac{7 \mksi^3+33 \mksi^2+113 \mksi-533}{96\ep}
-\frac{\mksi^2+4 \mksi+19}{16\ep}\,\zt\right)\,\CA^3\, a^3,
\\
Z_f^{\mathcal{N}=1}&=&
1
-\frac{\mksi}{\ep}\,\CA \,a
+\left(\frac{3\mksi (\mksi+1)}{4\ep^2}
-\frac{\mksi^2+8 \mksi+3}{8\ep}\right)\,\CA^2\, a^2
\nonumber\\
&+&\left(-\frac{\mksi \left(4 \mksi^2+9\mksi+9\right)}{8 \ep^3}
+\frac{\mksi^3+7\mksi^2+11 \mksi+3}{4\ep^2}
\right.\nonumber\\
&&\left.
\quad-\frac{10 \mksi^3+39 \mksi^2+159\mksi-66}{96\ep}
-\frac{\mksi^2+2\mksi+9}{8\ep}\,\zt\right)\, \CA^3\,a^3,
\\
Z_s^{\mathcal{N}=1}&=&
1
-\frac{\mksi-2}{\ep}\,\CA\,a
+\left(\frac{3\mksi^2-5 \mksi-4}{4 \ep^2}
-\frac{\mksi^2+8 \mksi-17}{8 \ep}\right)\,\CA^2 \,a^2
\nonumber\\
&+&\left(\frac{-12 \mksi^3+9 \mksi^2+33\mksi+32}{24 \ep^3}
+\frac{\mksi^3+6\mksi^2-7 \mksi-16}{4\ep^2}
\right.\nonumber\\
&&\left.
\quad-\frac{10 \mksi^3+39 \mksi^2+167\mksi-634}{96\ep}
-\frac{\mksi^2+11}{8\ep}\,\zt\right)\, \CA^3\, a^3,
\\
Z_{gh}^{\mathcal{N}=1}&=&
1
-\frac{\mksi-3}{4 \ep}\,\CA \,a
+\left(\frac{\mksi+21}{32 \ep}
+3\,\frac{\mksi^2-9}{32 \ep^2}\right)\, \CA^2\,a^2
\nonumber\\
&+&\left(\frac{-5\mksi^3-9 \mksi^2+9 \mksi+189}{128\ep^3}
+\frac{8 \mksi^3+39 \mksi^2+12\mksi-891}{384 \ep^2}
\right.\nonumber\\
&&\left.
\quad-\frac{3 \mksi^3+6 \mksi^2-139}{192\ep}
+\frac{\mksi^2+4 \mksi+19}{32\ep}\,\zt\right)\, \CA^3\,a^3.
\end{eqnarray}

\end{document}